\documentclass[12pt,a4paper]{article}
\usepackage{times}
\usepackage{graphicx}
\usepackage{placeins}
\usepackage[export]{adjustbox}
\usepackage{ifxetex}
\usepackage{svg}
\usepackage{amsthm}
\usepackage{subcaption}
\usepackage{pgfplots}
\usepackage{algorithm}
\usepackage{algpseudocode}
\pgfplotsset{compat=1.18}
\usepackage[margin=1in,footskip=0.5in]{geometry}
\usepackage{setspace}
\usepackage[]{amsmath}
\usepackage{amsfonts}
\usepackage{sectsty}
\usepackage[utf8]{inputenc}
\usepackage[round,semicolon,authoryear]{natbib}
\usepackage{xcolor}
\usepackage[space]{grffile} 
\usepackage{epstopdf}
\usepackage[space]{grffile}
\usepackage{xspace}
\usepackage{mathtools}
\usepackage[labelfont=bf,font=small]{caption}
\usepackage{multirow,array}
\usepackage{parskip}

\definecolor{blue1}{HTML}{2851CC}
\definecolor{red1}{HTML}{E00000}
\usepackage{hyperref}
\hypersetup{
    colorlinks,
    linkcolor = blue1,
    citecolor = blue1,
    urlcolor=blue1
}
\usepackage[capitalise,noabbrev]{cleveref} 
\crefformat{equation}{Eq.\,(#2#1#3)}
\Crefformat{equation}{Eq.\,(#2#1#3)}

\newtheorem*{proposition01}{Proposition (Singh et al., 2000)}

\newcommand\blfootnote[1]{%
  \begingroup
  \renewcommand\thefootnote{}\footnote{#1}%
  \addtocounter{footnote}{-1}%
  \endgroup
}

\usepackage[normalem]{ulem}

\newcounter{marginNoteCounter}
\newcommand{\ie}   			{i.e.\@\xspace}
\newcommand{\eg}   			{e.g.\@\xspace}

\newcommand{\pN} {p^\text{N}}
\newcommand{\pM} {p^\text{M}}

\newcommand{\pA} {p_\text{A}}
\newcommand{\pB} {p_\text{B}}
\newcommand{\PrcSet} {P}

\begin{document}

\title{Algorithmic collusion under asynchronous price updating}
\date{}
\author{Ivan Conjeaud$^\dagger$ $^\diamond$ \quad Gaspard Abel $^{\ddagger} $$^*$ \quad Argyris Kalogeratos$^\ddagger$}

\maketitle
\begin{abstract}
    This paper investigates the effect of asynchrony in agents' updates in the emergence of algorithmic collusion. We present a continuous-time model for algorithmic collusion in which two firms use $Q$-learning algorithms to set  prices asynchronously in a Bertrand duopoly. The firms update their prices at times dictated by a Poisson clock.	By controlling the extent of agents' asynchrony, we run extensive numerical experiments with three specifications of the algorithm to investigate the emergence of algorithmic collusion. The strength of collusion is measured by a standard collusion index, as well as by  automatically detecting the reward-punishment schemes. This is done by recording a large number of algorithms' reactions to unilateral price cuts and comparing them with the reactions of untrained algorithms. Our findings indicate that asynchrony hampers collusion, especially when the algorithms are stateless. When they condition on their competitor's previous prices, the sensitivity of algorithmic collusion to asynchrony varies depending on the type of information they have access to.  The implications of these results for the regulation of algorithmic pricing are discussed.
    \medbreak
\end{abstract}
\blfootnote{\textit{Corresponding author}: \href{mailto:ivan.conjeaud@psemail.eu}{ivan.conjeaud@psemail.eu}}
\blfootnote{$\dagger$: \textit{Paris
School of Economics, Université Paris 1 Panthéon Sorbonne}, 48 Blvd. Jourdan, 75014 Paris, France.}

\blfootnote{$\diamond$: \textit{Aix Marseille Université, CNRS, AMSE} Marseille, France.}

\blfootnote{$\ddagger$: \textit{Centre Borelli, ENS Paris Saclay}, 4 Av. des Sciences, 91190 Gif-sur-Yvette, France.}
\blfootnote{$^*$: \textit{CAMS, EHESS} 54 Blvd. Raspail, 75006 Paris, France.}
\newpage
\section{Introduction}
It is a well-established fact that automated pricing is now prevalent in many sectors \citep{adams2025rise, chen2016empirical}. This has been a source of concern for regulators \citep{algorithms2017collusion, calvano2020protecting}, especially after the seminal work of \citet{waltman2008q} and the influential numerical experiments of \citet{calvano2020artificial}. Their results have highlighted how reinforcement learning algorithms learn collusive strategies without being instructed to do so. Existing algorithmic collusion models assume that all algorithms update their price at the same moment or sequentially. However, unless enforced, such synchrony is unlikely and motivates the development of pricing algorithms that update their prices infrequently. In this paper, we present a model of algorithmic collusion for the setting in which the algorithms do not necessarily update their price simultaneously, and show that \textit{asynchrony} hampers the emergence of algorithmic collusion.

Automatic agents operating in the same market may perform pricing updates, not only at different times, but also at different frequencies due to differences in technology and availability of computational resources or information acquisition about the market environment.
It is practically infeasible to adjust prices in real time 
as they are time-consuming \citep{koushik2012retail, pekgun2013carlson}, and they usually are detrimental to the firm due to negative consumer perception \citep{haws2006dynamic}. 
In practice, pricing algorithms are often triggered by events. For instance, a common reason to use dynamic pricing is to manage an inventory, especially for perishable goods \citep{elmaghraby2003dynamic}. Variations in the inventory of a firm are typically due to surge or decline in consumers demand, or proxies such as visits to the firm's website, which might or might not be correlated across firms. Some pricing solutions developed 
by specialized providers even include the possibility to define triggers 
in the form of rules\footnote{See for instance \href{https://repricer.com}{https://repricer.com}}. 
Updating prices at random times has a direct advantage as a protective mechanism, since it prevents competitors or consumers from predicting when price drops are likely; the literature on random discounting has made that effect on consumers clear \citep{dilme2025dynamic, chen2023intertemporal}. 
On the competitors' side, the timing of price updates can also be a point of deliberate exploitation for rivals. 
For instance, the website repricer.com, a dynamic pricing service provider, 
gives the following piece of advice for pricing scheduling: "\emph{Most competitors adjust prices during business hours. Schedule your most aggressive moves during their off-hours to capture maximum advantage.}" \footnote{Source (accessed in July 2026): \href{https://repricer.com/blog/repricer-scheduling-feature-guide/}{https://repricer.com/blog/repricer-scheduling-feature-guide/}}. The possibility that time periods of idleness could be exploited by a rival strategy makes clear why random price updates might also  strategic reasons for protection. 

Investigating the effects of asynchrony on the collusive behavior of algorithms is also relevant for regulatory practices. An 
example is the long-lasting debate concerning \emph{FuelWatch}, in Western Australia. Since 2001, the government of Western Australia requires, by law, that the firms operating petrol stations update their prices simultaneously once a day to increase market's transparency for consumers. Since then, the effect of this measure on prices has been a source of intense debate between the Australian Competition and Consumer Commission and scholars \citep{davidson2008secret, harding2008foolwatch}, although the academic consensus points to evidence of increased prices and tacit collusion 
\citep{wang2009mixed, byrne2019learning, adams2025impact}. In contrast, in Germany, the \textit{Bundeskartellamt} has implemented a policy to increase market transparency for consumers -- the \textit{Markttransparenzstelle für Kraftstoffe} (MTS-K) -- without requiring retailers to update their prices simultaneously, which led to pro-competitive results \citep{montag2026does}. For this sector, which has seen algorithmic pricing increasing the profit margins \citep{assad2024algorithmic}, investigating the effects of synchrony on algorithmic collusion is thus of primary importance. 

Previous contributions investigating this market have studied how market transparency affects tacit collusion. In this work, we highlight a different perspective and study the intrinsic effects of algorithmic synchrony on the behavior of learning algorithms. 
Taking this specific market and FuelWatch as an example, our results provide another argument against such initiatives: as we show, combining synchrony and market transparency increases the risk of algorithmic collusion.  

In stark contrast to such properties
, the literature on algorithmic collusion has focused on discrete-time models 
where algorithms update simultaneously 
\citet{calvano2020artificial} or sequentially 
\citet{klein2021autonomous}. The present article aims to fill this gap in the literature and answer the following question: how is algorithmic collusion affected when the algorithms' updates are asynchronous?

We introduce 
a model of a duopoly \textit{à la }Bertrand, in which two firms employ $Q$-learning algorithms for their pricing updates. Unlike previous contributions on algorithmic collusion that use deterministic discrete-time updates, we assume that each algorithm has its own continuous-time clock, namely a Poisson process, dictating its next update time, 
while the opponent algorithm can as well perform an update at the same time with probability $q \in [0,1]$. The parameter $q$ controls the synchrony of updates, from being completely independent Poisson clocks with different rates ($q=0$), to perfectly synchronous updates dictated by a single clock ($q=1$). We study the effect of $q$ on algorithmic collusion under three different specifications for the algorithms. First, we investigate a setting with stateless algorithms \textit{à la} \citet{banchio2023adaptive}. We find that this type of collusion is strongly affected by asynchrony, and hence does not appear for low values of $q$. Asynchrony directly affects the mechanism responsible for collusion by stateless algorithms as it prevents them to efficiently synchronize on mutually beneficial prices. Then, we consider two variations of $Q$-learning algorithms that condition on their opponent's price: when they condition on the average price set by their opponent since their own last update, and when they condition on the current price of their opponent. These two variations correspond to two levels of market transparency: in the first case, the algorithms only have access to aggregate information about their opponent's behavior, while in the second one they have access to their opponent's current price, through web-scraping for instance. We run extensive numerical simulations and use two metrics to quantify algorithmic collusion. First, a standard collusion index measuring how much larger are the joint payoffs compared to the competitive (Nash) payoffs. Second, we detect the presence of reward-punishment schemes learned by the algorithms in the long run. We use a method, novel in the context of algorithmic collusion, that allows us to quantify their emergence by automatically detecting their pattern: we record the reaction of the algorithms to unilateral price cuts by their opponent at the end of the $T$ iterations. We do the same exercise with $Q$-learning algorithms \textit{prior} to them interacting, so that their behavior is random and driven by their initial conditions. We then perform a pattern-detection exercise: we measure the similarity between the algorithms' reactions to price cuts using the Dynamic Time Warping (DTW) distance, and we also use that to perform clustering with the DBSCAN algorithm to infer the structure of the interaction process. We find that increasing $q$ leads to increased collusion in the two $Q$-learning setups, as both the similarity metric and the clustering-based metric reveal. A point of differentiation, though, is that when the algorithms condition their action on the current price of their opponent, we find that collusion is very robust to asynchrony, hence persists despite varying $q$, while in the second case high values of $q$ are needed for collusion to emerge.

These results have important implications for the regulation of algorithmic pricing:

\begin{itemize}
\item First, they suggest that the exogenous addition of noise in the update times of pricing algorithms can mitigate algorithmic collusion. 
\item Second, it shows that algorithmic collusion is more likely when algorithms can detect and react to the current price update of their competitors. 
\item Finally, they suggest that the concerns over algorithmic collusion might be exaggerated in many situations where there is no easy way to induce correlation between one's own update and their opponent and where the algorithms do not have an easy access to their competitor's price. 
\end{itemize}

The rest of the paper is organized as follows. \Cref{sec: lit} reviews the related literature
. \Cref{sec: model} presents the proposed model. \Cref{sec: stateless} presents our results on stateless algorithms. \Cref{sec: method} presents our experimental setup and the numerical methods we use for the algorithms with states. \Cref{sec: results} presents the results, \Cref{sec:discussion} discusses their implications, and finally \Cref{sec: conclusion} concludes the study.
 
\section{Related literature} \label{sec: lit}
This article contributes to the growing literature on algorithmic collusion. Since the early contribution of \citet{waltman2008q}, and more so since the influential paper of \citet{calvano2020artificial}, a lot of effort has been devoted to investigating when algorithmic collusion appears. Extensions of the original setting, including models with more advanced algorithms \citep{hettich2021algorithmic}, imperfect monitoring \citep{calvano2021algorithmic}, and different economic environments \citep{ballestero2026algorithmic, ye2025algorithmic}, have pointed toward the robustness of this phenomenon. Despite the rich variety of settings in which algorithmic collusion has been investigated, the timing assumptions have been limited to either completely synchronous updates, or sequential ones. This paper fills this gap in the literature and shows that the common assumption on timing is not innocuous.

The exact nature and definition of ``algorithmic collusion'' is a topic of debate, even more so that it includes at least three different but related phenomena. First, it has been long documented that misspecified duopolists estimating a demand function as if they were monopolists, tend to set supracompetitive prices. Articles investigating this phenomenon include \citet{cooper2015learning}, \citet{hansen2021frontiers}, \citet{cho2024collusive} and \citet{baek2026misspecified}. However, the one that has received most attention is the tendency of $Q$-learning algorithms to learn a form of \textit{tacit collusion}, where \textit{collusion} here is understood as the fact that a firm \textit{causes the other one to set supracompetitive prices} \citep{harrington2018developing} by means of reward-punishment schemes. The models 
highlighting this tendency typically assume 
algorithms that are able to condition on their opponent's past pricing in order to implement reward-punishment strategies, and include all of the aforementioned contributions. A second one has been focusing on the behavior of \textit{stateless} algorithms, which by definition, cannot directly enforce tacit collusion. When the $Q$-learning algorithms do not have a memory, they can nevertheless engage in \textit{spurious} collusion, as coined by \cite{calvano2023algorithmic}. \citet{banchio2023adaptive} 
formally uncovered the \textit{spontaneous coupling} phenomenon that causes the algorithms to alternate phases of defection and cooperation when they repeatedly play a prisoner's dilemma. Before them, \citet{banchio2022artificial} and \citet{asker2022artificial} had observed this phenomenon in different contexts, and in a recent contribution \citet{douglas2026illusion} study a similar setting with bandit algorithms. These contributions highlight the importance of three elements for the emergence of spurious collusion: (1) a low exploration rate; (2) the fact that only selected actions are updated \footnote{\citet{asker2022artificial} name this property \textit{asynchronous updating}. We use this term in a very different, and, we believe, much better adapted way.}; (3) the frequency of symmetric profiles of actions played by the algorithms\footnote{\citet{douglas2026illusion} use the term ``synchrony'' for this property. To avoid confusion, we will not use this term and reserve it for when the algorithms update \textit{at the same time}.} (\citet{douglas2026illusion}). These three properties are essential for the algorithms to \textit{synchronize} on mutually beneficial actions. In this paper, we highlight a fourth essential property needed for spurious collusion to happen: the fact that the algorithms update their actions at the same time.

There has been also questioning regarding the robustness and the likelihood of algorithmic collusion, or even the relevance of the concept, even when the algorithms learn strategies that resemble reward-punishment schemes. \citet{epivent2024algorithmic} note that the algorithms punish price surges as well as price cuts, even when the algorithms learn under-competitive prices. \citet{abada2023artificial} and \citet{lambin2024less} claim that ``seemingly collusive outcomes'' are due to a lack of exploration, while \citet{abada2024collusion} also note that more advanced algorithms outperform $Q$-learning while not colluding. A similar point is made by \citet{carissimo2025algorithmic}. Finally, \citet{eschenbaum2022robust} show that $Q$-learning algorithms do not practice collusive pricing when they are trained outside of their operating environments. 

In this paper, we take a different stance and claim that algorithmic collusion is also dependent on a strong assumption about the synchrony of updates, which is unrealistic in most markets, and on the precise observability of the competitor's current price, which is not obvious. This fact affects both spurious and genuine (or, following Lambin, ``seemingly'') collusive outcomes. While the first one does not require any observation of the competitor's price, it disappears when updates are not synchronous. The second one is more robust to asynchrony, granted that precise observations of the competitor's price can be made.


The present article also provides elements relevant to the literature on the regulation of algorithmic collusion. Early contributions have highlighted that the possibility of tacit collusion by autonomous algorithms does not contravene to the Sherman Act and provided avenues for regulation practices. While \citet{harrington2018developing} advocates for examinations of the code used by pricing algorithms as well as real-life tests to monitor their output, \citet{hartline2024regulation, hartline2025regulation} take a different approach and propose an output based statistical test to detect collusion. On the other hand, \citet{mehra2016antitrust} makes the case for proactive rather than reactive regulation, and the need to establish good practices and guidelines. \citet{ezrachi2017artificial} provide a typology of algorithmic collusion and use it to formulate avenues for regulation. In particular, they advocate for action on market transparency and for technical restrictions on the data used by the algorithms. The results of the present article point in this direction as well, since under the more realistic assumption of update asynchrony, collusive outcomes depend strongly on what the algorithms use to condition their price on.

\section{The model} \label{sec: model}
\subsection{Economic environment}
Two firms, A and B, sell a differentiated product on the same market and compete on prices. Time is continuous and denoted by $t \in \mathbb{R}_+$. We assume a standard logistic demand. For $i \in \{\text{A}, \text{B}\}$, if prices $(p_i, p_{-i})$ are set by the two firms
, then the instantaneous rate of demand to firm $i$ 
writes:
\begin{equation}\label{eq:demand}
    D_i(p_i,p_{-i})
		=\frac{\exp\big({\frac{a-p_i}{\mu}}\big)}{\exp\big({\frac{a-p_i}{\mu}}\big) + \exp\big({\frac{a-p_{-i}}{\mu}}\big)+1}
		.
\end{equation}
The parameter $a>0$ captures vertical differentiation (\ie  measurable differences between the two products, like different qualities), while $\mu>0$ captures horizontal differentiation (subjective differences between the products). Given the per unit cost $c>0$ of production, the instantaneous rate of profit realized by firm $i$ during this time is: 
\begin{equation}\label{eq:profit}
\pi(p_i,p_{-i})
= D(p_i,p_{-i})(p_i-c)
.\end{equation}

The monopoly price in this market, defined as the symmetric price that maximizes joint profit, is denoted by $\pM$. 
In a static Bertrand duopoly with logistic demand functions, there is a unique Nash equilibrium, which we denote 
by the price $\pN$. 
The \textit{collusion index} of a pair of prices $(\pA, \pB)$ is 
defined as:
\begin{equation}\label{eq:collusion-index}
    CI(\pA, \pB)=\frac{
    \frac{1}{2}\big(\pi(\pA,\pB)+\pi(\pB,\pA)\big) - \pi(\pN,\pN)}{\pi(\pM, \pM)- \pi(\pN, \pN)}.
\end{equation}
This quantity is equal to $1$ when both firms charge the monopoly price ($\pA = \pB = \pM$), and equal to $0$ when they charge the competitive price ($\pA = \pB = \pN$).

\begin{figure}
    \centering
    \includegraphics[width=0.7\linewidth]{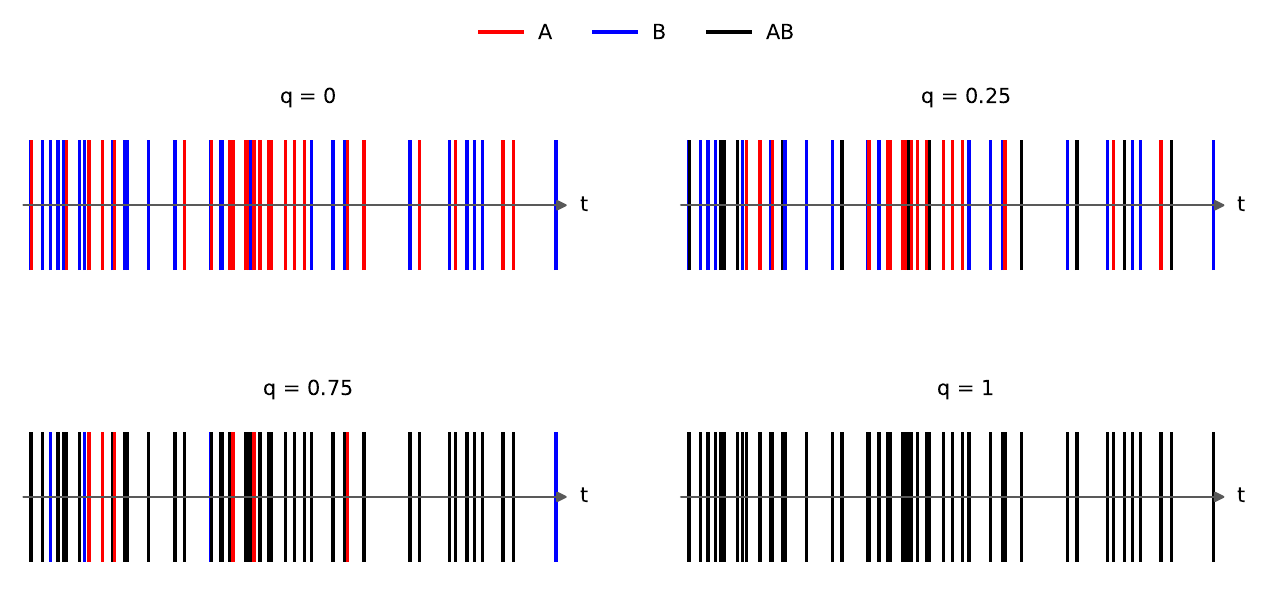}
    \caption{\textbf{Illustration of the timing of updates} for different values of $q \in [0,1]$, which controls the updating synchrony from independent to totally synchronized updates.}
    \label{fig: illustration}
\end{figure}

The firms can revise their price at \emph{exogenous} update times; let the countable set of update times for firm $i$ be denoted by $T_i=\{t^i_k \,|\, k \in \mathbb{N}, t^i_0=0\}$. We assume that each player's update times are dictated by independent Poisson clocks of parameter $\frac{\lambda}{2}$, and additionally, when one of the players updates their price, the other player updates theirs with probability $q \in [0,1]$. When $q=1$, the model is equivalent to both players always updating simultaneously at update times dictated by a single Poisson clock of parameter $\lambda$. When $q=0$, it is equivalent to a model in which the players update their price at times dictated by two independent Poisson clocks of parameter $\frac{\lambda}{2}$. \Cref{fig: illustration} illustrates the timing of the algorithms' updates for $\lambda = 1$ when $q$ varies between $0$ and $1$: in red, the times at which $A$ updates its price, in blue, the times at which $B$ updates its price, in black, the times at which both update their price. Mathematically, this assumption on the timing of updates has at least two other interpretations, that we detail in Appendix \ref{sec: timing}.

\subsection{$Q$-learning in continuous-time}
We assume that the firms employ $Q$-learning algorithms to select their prices from a set of $n \ge 2$ possible prices $\PrcSet$, 
equally spaced in the interval $[\pN- \xi, \pM + \xi]$, where $\xi>0$ is a parameter. 
%
A description of regular $Q$-learning is included in \Cref{sec: $Q$-learning} for reference. As with regular $Q$-learning, each algorithm keeps in memory and updates a matrix of values, called the $Q$-matrix, which is given by $Q : \PrcSet \times \PrcSet^2 \rightarrow \mathbb{R}$. We denote by $Q_t(p,s)$ the $Q$-value associated with price $p$ and state $s$ at time $t$. The state is an ordered pair of prices whose value depends on the information that the algorithm has access to, which can then be used to condition its pricing. 

For $k \in \mathbb{N}^*$, consider the time at which a firm updates for the $k$-th time, and denote by $p_{t_{k-1}}$ the price that was set 
by its previous update, and by $s_{t_{k-1}}$ the last state. 
The 
update rule for 
the $k$-th update of the $Q$-matrix is given by:
\begin{equation}
\begin{cases}
Q_{t_k}(p, 
s_{t_{k-1}})=(1-\alpha)Q_{t_{k-1}}(p, s_{t_{k-1}})
  + \alpha\Big[ \overline{\pi} + \gamma \,{\displaystyle \max_{p' \in \PrcSet}} \,Q_{t_{k-1}}(p', 
	s_{t_k}) \Big],
  & \text{if } p = p_{t_{k-1}} \\
Q_{t_k}(p, s)=Q_{t_{k-1}}(p, s), & \text{otherwise}.
\end{cases}
\end{equation}
This updates the $Q$-value for the chosen price and the state $s_{t_{k-1}}$, while keeping frozen the rest of the $Q$-matrix. The parameter $\alpha \in [0,1]$ is  referred to as the \emph{learning rate} and controls  the weight put on new information in the updating, while $\gamma \in [0,1]$ is a discount rate controlling  how much weight is put on future payoffs.

We consider the $Q$-learning case with no state (\textsc{NoState}), as well as two variations for when a state is used. In the \textsc{NoState} case, the $Q$-matrix reduces to a vector, thus we can simply write $Q_t(p) := Q_t(p,\text{null})$. In the first of the specifications with state, \textsc{AvgPrice}, let $\overline{p}_{t_{k-1}}$ be the price charged on average by its opponent 
during the period $[t_{k-1}, t_{k})$ and $\hat{p}_{t_{k-1}}=\arg \min_{p \in \PrcSet }|\overline{p}_{t_{k-1}}-p|$ the price in the grid closest to $\overline{p}_{t_{k-1}}$. The state $s_{t_k}$ is the couple $(p_{t_{k-1}},\hat{p}_{t_{k-1}})$. In the second one, \textsc{CurrentPrice}, the algorithms use the current price of the opponent so that $s_{t_{k}} =(p_{t_{k-1}},p'_{t_k})$.
The payoff used 
is the average payoff $\overline{\pi}$ collected between update times $t_{k-1}$ and $t_k$. Note that when $T_\text{A}=T_\text{B}$ the update rule is equivalent to the one used in \citet{calvano2020artificial} in the two specifications with states, while it is equivalent to the one used in \citet{banchio2023adaptive} in the stateless case.

When updating at time $t$, the algorithms apply a $\varepsilon$-greedy policy. Following \citet{banchio2023adaptive}, we consider a constant exploration rate in the stateless environment. Otherwise, when the algorithms condition on their opponent's past prices, we use an exploration rate, denoted by $\varepsilon_t$, with an exponential decay:
\begin{equation}
    \varepsilon_t=\exp(-\beta t), \ \ \ \beta >0.
\end{equation}
At state $s$, the price is selected according to:
\begin{equation}
    \begin{cases}
        p=\arg \max_{p'\in \PrcSet} Q_{t}(p', s), & \text{ with probability }  1-\varepsilon_t \\ p \sim \mathcal{U}(\PrcSet), & \text{ with probability } \varepsilon_t.
    \end{cases}
\end{equation}
The parameter $\beta$ controls how fast the exploration rate vanishes. For $\beta=0$, the algorithms always explore uniformly options from $P$, while for $\beta=+\infty$ the algorithms are 
greedy and select the price with highest $Q$-value.

\section{Stateless $Q$-learning algorithms}
\label{sec: stateless}
In this section, we consider stateless $Q$-learning algorithms with constant exploration rate as in \citet{banchio2023adaptive}. We simulate the $Q$-learning algorithms for a Bertrand duopoly with $a=2$, $\mu=0.25$, $c=1$, $n=20$, and $\xi=0.1$, for different fixed values of $\varepsilon$ and $q$. We measure the limiting average payoffs ($100$ independent runs) and compute the corresponding collusion index (CI, \Cref{eq:collusion-index}). We also compute the collusion index induced by the discretization of the action space. Since the action space is discretized, the Bertrand duopoly game with the discrete action space might feature several pure strategy Nash equilibria, evaluated by different CI values. In our experiments, the maximal collusion index pure strategy Nash equilibria induced by the discretization of the action spaces is $\text{CI}=0.043$, so that any pair of parameters $(q,\varepsilon)$ under which the measured collusion level $\text{CI} < 0.043$ can be considered as non-collusive. 
\begin{figure}[t!]
    \centering
    \includegraphics[width=0.8\linewidth]{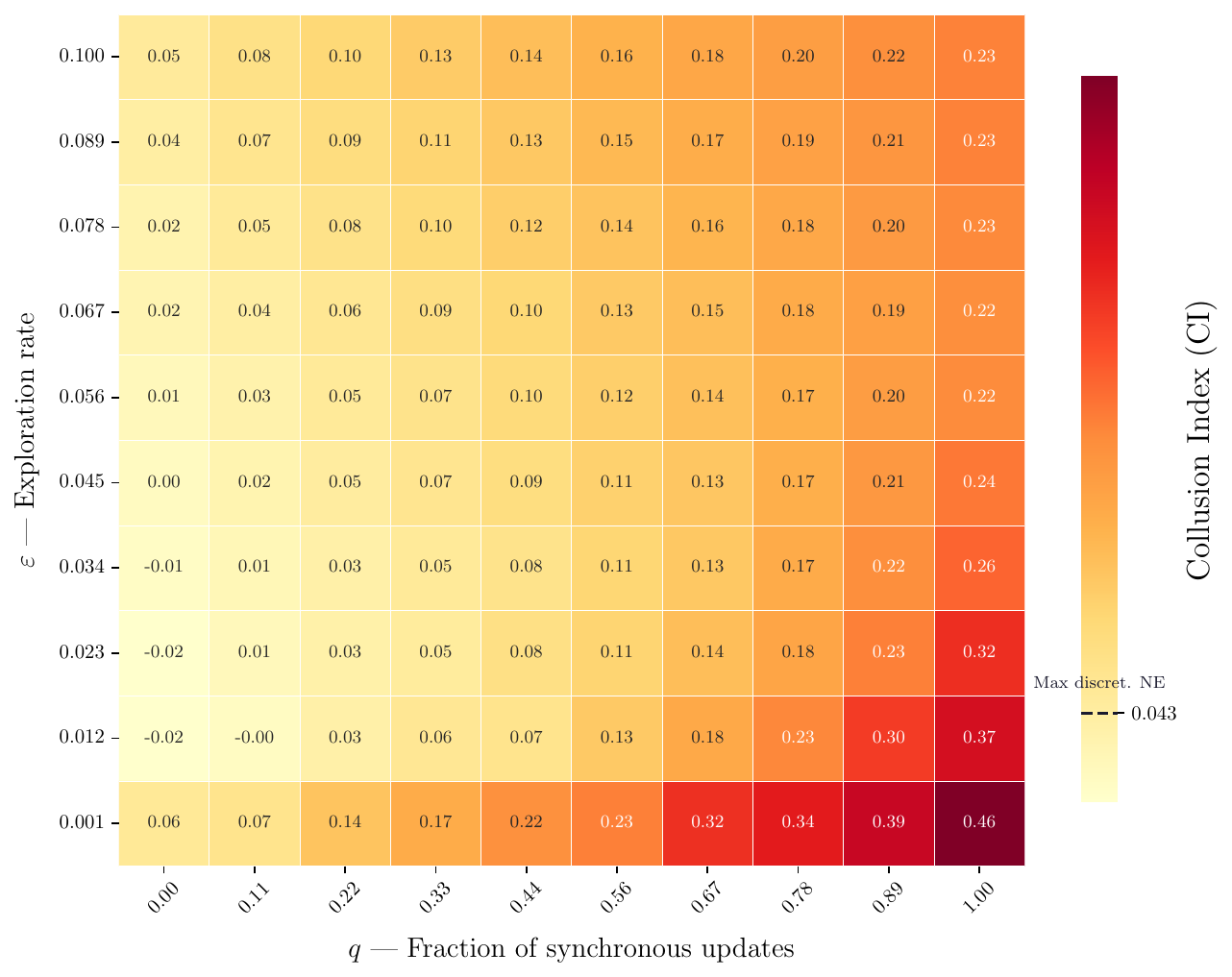}
    \caption{\textbf{Collusion indices with stateless algorithms} with $n=20$ and $\alpha=0.1$. The collusion index is lower for high exploration rates and low synchrony, and comparable to the competitive outcome for $q=0$.}
    \label{fig: hm stateless}
\end{figure}

The results of these experiments are shown in \Cref{fig: hm stateless}. In the figure, when fixing $\varepsilon$, higher collusion appears when $q$ gets larger (\ie along the rows of the heatmap), and hardly any collusion 
for low enough values of $q$. However, the effect of $\varepsilon$ is non-monotonic: this is due to two effects. When $\varepsilon$ increases, the algorithms tend to learn less collusive strategies, as has already been known from \citet{banchio2023adaptive}, so that the collusion index is first decreasing with $\varepsilon$. However, for  large enough $\varepsilon$ values, the algorithms do not learn collusive strategies, but mechanically engage in more exploration and choose higher prices more often, which causes the collusion index to increase.
Asynchrony directly affects the mechanism through which the algorithms coordinate on high prices. In order to understand how this takes place, we consider the simplest case, with two prices: high and low. In this case, the stage game played by the algorithms is a prisoner's dilemma where setting a high price corresponds to Cooperation (C) and setting a low price corresponds to Defection (D). As has already been observed in previous works \citep{banchio2023adaptive, asker2022artificial}, under synchrony the algorithms alternate between cooperative and competitive phases. When the algorithms both set high prices, they reinforce each other into setting high prices by making each other's $Q$-values for high prices increase. When by exploring they choose a low price, they quickly realize that setting low prices is better. But as they do so, their $Q$-values for low prices tend to \textit{decrease}. Since they only update the $Q$-values associated to the actions they choose, their $Q$-values for low prices end up undercutting the ones for high prices \textit{at the same time}: when they do so, they start setting high prices again, and a new cycle begins. By contrast, when the updates are asynchronous, if the two algorithms have set low prices, the one who updates can indeed switch to a high price as the $Q$-value for low price decreases. But since the opponent  algorithm does not update its price at the same time, it keeps its price low, which prevents the two algorithms from cooperating again. 

\begin{figure}
    \centering
    \includegraphics[width=0.8\linewidth]{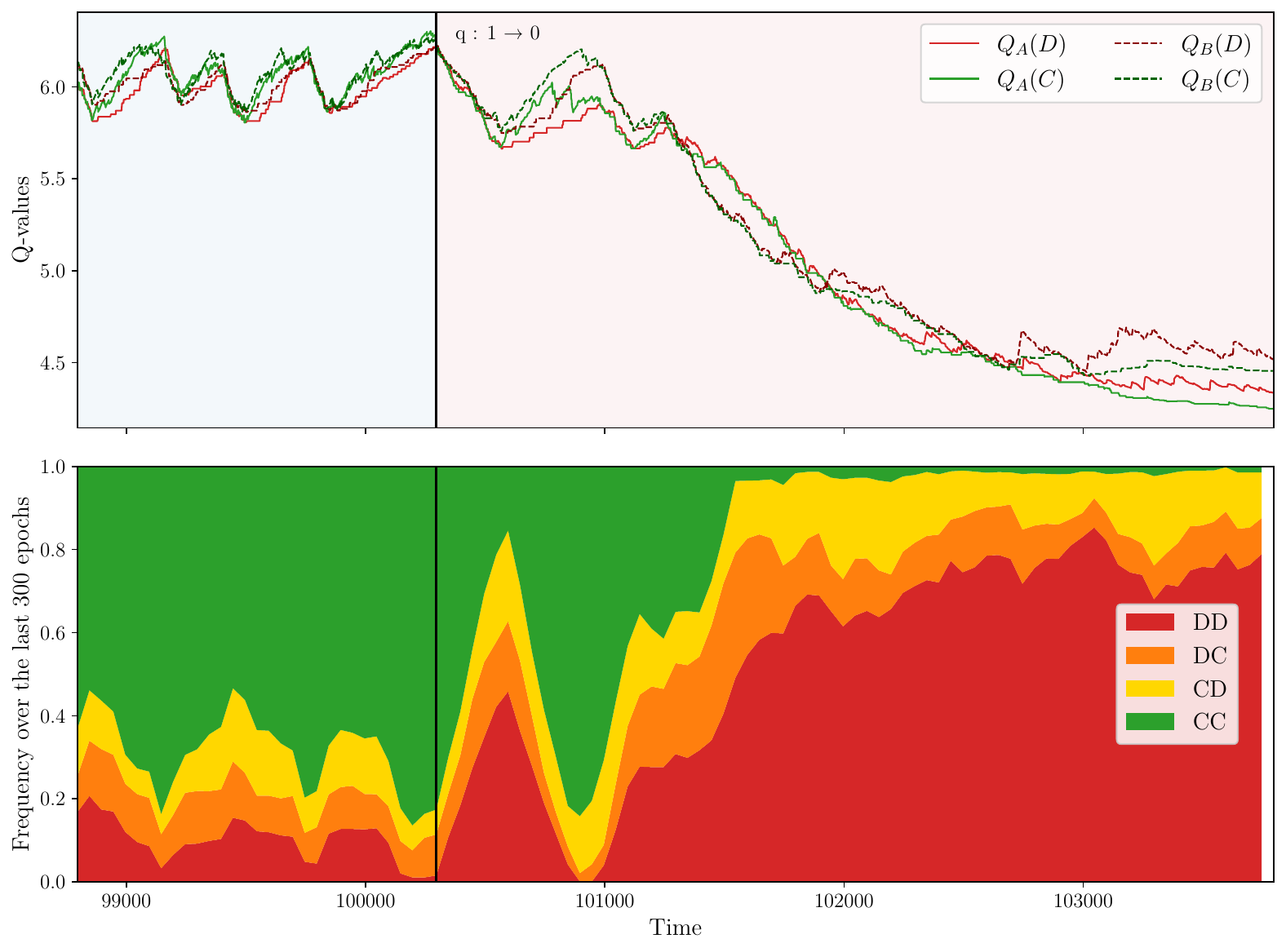}
    \caption{\textbf{Switching regime experiment.} We start the experiment with full synchrony ($q=1$) and exogenously change it to full asynchrony ($q=0$; the time is indicated by a black vertical line). Following the change, the algorithms fail to synchronize on mutually beneficial actions.}
    \label{fig: switch experiment}
\end{figure}
To support this explanation, we design the following experiment. We run the stateless $Q$-learning algorithms with two prices and the same parameterization for $T=10^5$ synchronous  updates ($q=1$), then we change to completely asynchronous updates ($q=0$). For producing a clearer visual, we set a high exploration rate, $\varepsilon = 0.2$, in order to accelerate the switching behavior. We keep track of the $Q$-values and the actions selected by the algorithms ($C$ for high price, $D$ for low price). The results for one representative instance are presented in \Cref{fig: switch experiment}. The top panel shows the evolution of $Q$-values around the switching point, while the bottom panel displays the fraction of each profile over the interval $[t-W, t]$ of %
time, where $W=300$. Before the change of regime, the alternation of cooperation and defection is visible: the $Q$-values tend to increase in cooperative phases, then they decrease, then increase again when the algorithms come back to cooperation. The high frequency of cooperation causes the $Q$-values to remain high. After the regime change, the $Q$-values inexorably decrease, which attests that the algorithms fail to synchronize back on mutual cooperation. Eventually, they end up in a regime with low $Q$-values, where they defect most of the time, which in this context means that they do not collude over high prices.

\section{$Q$-learning algorithms with states}
\label{sec: method}
When algorithms condition on each other's price, algorithmic collusion is characterized by supra-competitive profits collected by the algorithms in the long run 
and a form of reward-punishment scheme is learned by the algorithms. Our aim is to use sensible metrics to capture both these phenomena, and hence measure the effect of update synchrony 
on algorithmic collusion. 

Our baseline parameterization is the same as the one in \citet{calvano2020artificial}: $a=2, \mu=0.25, c=1, n=15$ and $\xi=0.1$. The Nash and monopoly prices are found computationally by solving the associated first order conditions using a simple bisection method. For the learning parameters, we set $\alpha=0.1$, $\beta=10^{-5}$ and $\gamma=0.95$, which makes exploration fade out exponentially in time. In each of our 
experiments, we run the algorithms so that each is updated for $T=10^6$ periods and adapt the number of total updates accordingly.

\subsection{Response graph}
As the exploration rate vanishes when $t$ grows large, the algorithms become greedy in the limit and at each update they only choose the actions with the highest $Q$-values. As a consequence, the strategies learned by the algorithms in the limit and their behavior can be represented by a digraph, which we call the \textit{response graph}.

The response graph represents each state $s$ as a node and thus comprises $|P|^2$ nodes. An outgoing edge is drawn from state $s=(\pA, \pB)$ to another state $s'$ in any of the three following cases. Either $s'=(\pA', \pB)$ and the $Q$-matrix of $A$ prescribes to choose price $\pA'$ in state $s$, or $s'=(\pA, \pB')$ and the $Q$-matrix of $B$ prescribes to choose price $\pB'$ in state $s$, or $s=(\pA', \pB')$ and the $Q$-matrix of $A$ prescribes to choose $\pA'$ in $s$ while the $Q$-matrix of $B$ prescribes to choose $\pB'$ in $s$. When $q=1$, edges are only drawn in the latter case as updates are always synchronous, so that each node has at most one outgoing edge. When $q=0$, edges are only drawn in the first two cases since updates are almost never synchronous.

\begin{figure}
    \centering
    \includegraphics[width=0.6\linewidth]{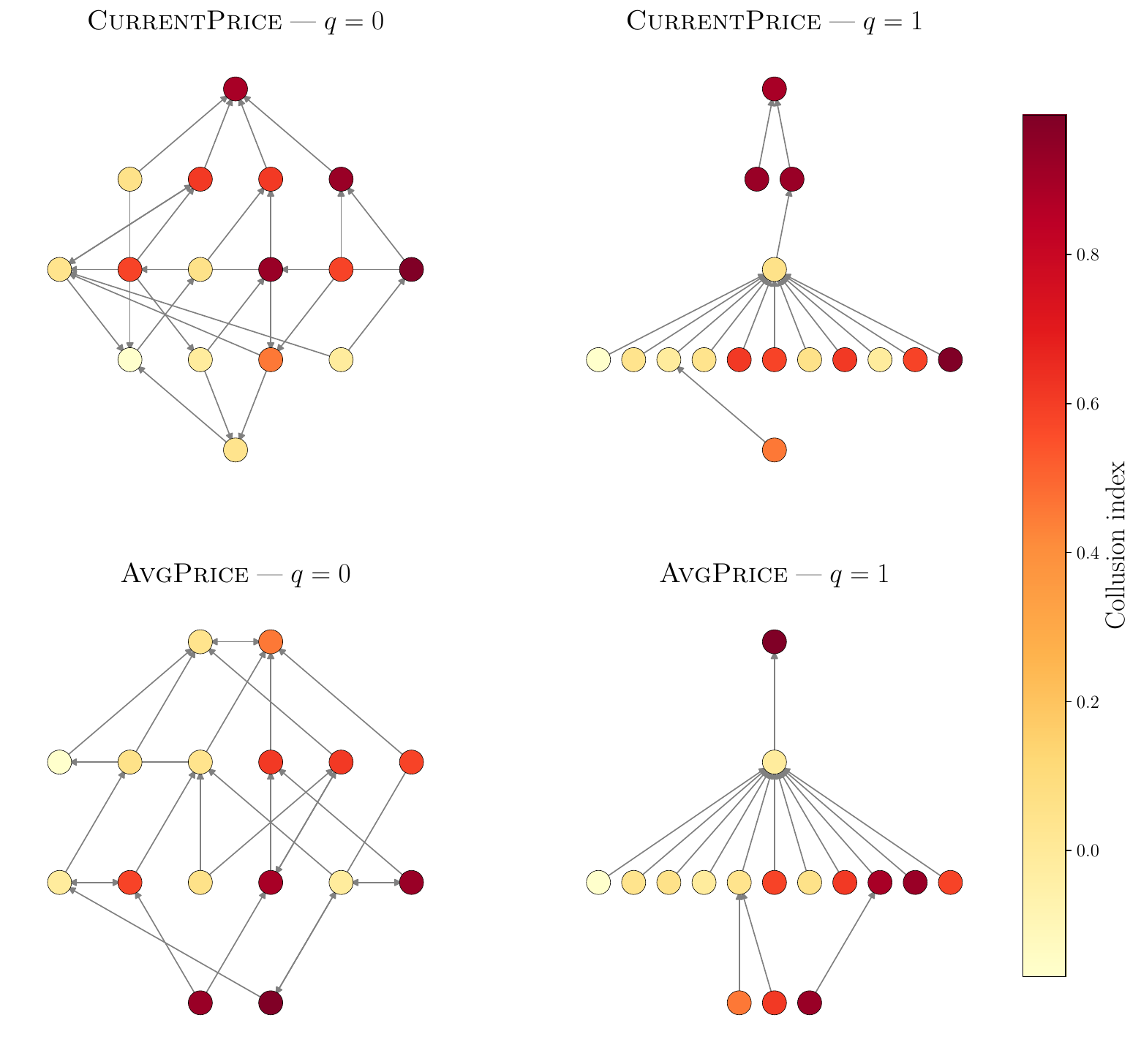}
    \caption{\textbf{Response graphs for different values of $q$ in the two specifications.} The color of the nodes represents the collusion index of the associated pair of prices. While \textsc{AvgPrice} for $q=1$ and \textsc{CurrentPrice} show structure indicating reward-punishment schemes, it is not the case for \textsc{AvgPrice} $q=0$.}
    \label{fig:response_graphs}
\end{figure}

\Cref{fig:response_graphs} compares the response graphs after four different runs of the algorithms with $n=4$ prices: one for $q=0$ and one for $q=1$ in each of the algorithm specifications. 
The colors of the nodes correspond to the collusion index of the pair of prices.

The response graphs reveal several important features. When $q=1$, the 
graphs are very similar to the one shown in \citet{calvano2020artificial}. There is a well-identified steady-state which is the unique sink of the response graph, with a high collusion index. The distances in the graph are short, suggesting quick returns to the steady state if a price cut occurs. Crucially, the presence of a central node with a very low collusion index reveals that the algorithms learn a reward-punishment scheme: a deviation from the steady-state is almost systematically punished by a price cut, followed by a return to the steady state. When the algorithms condition on the current price of their opponent with $q=0$, a similar structure appears, even though it is less visible due to each node having two outgoing edges. When $q=0$ and the algorithms condition on their opponent's average price since their own last update, this structure doesn't appear. There are two sinks, with low levels of collusion, no obvious central node, and distances that are longer. This suggests that in this last case, collusion is harder to sustain.

\subsection{Reaction to price cuts}

A defining feature of algorithmic collusion is precisely the presence of reward-punishment schemes in the algorithms' learned behavior: price cuts are punished, then both algorithms set high prices again. As is standard in the literature on algorithmic collusion, we observe reward-punishment patterns by recording the behavior of the algorithms following an exogenously imposed price cut. 

We propose a novel method to precisely measure this phenomenon so as to quantify its dependence on $q$.


For each run of the algorithms, we obtain the response graph and identify its sinks or cycles with no outgoing edges. We then perform random price cuts starting at the sinks, and record the trajectory of prices followed by the algorithms. In order to measure similarity between the trajectory of prices following price cuts, we use the Dynamic Time Warping (DTW) algorithm~\cite{berndt1994using} and the associated metric. DTW is a standard method to measure similarity between time series that allows one to detect patterns despite lags and deformations. It is a standard tool for pattern recognition, ranging from speech recognition and motion analysis to bio-signals and pattern recognition in financial markets \citep{keogh2005exact}. 

\begin{figure}[t]
    \centering

    \begin{subfigure}[b]{0.325\textwidth}
        \centering
        \includegraphics[width=\linewidth]{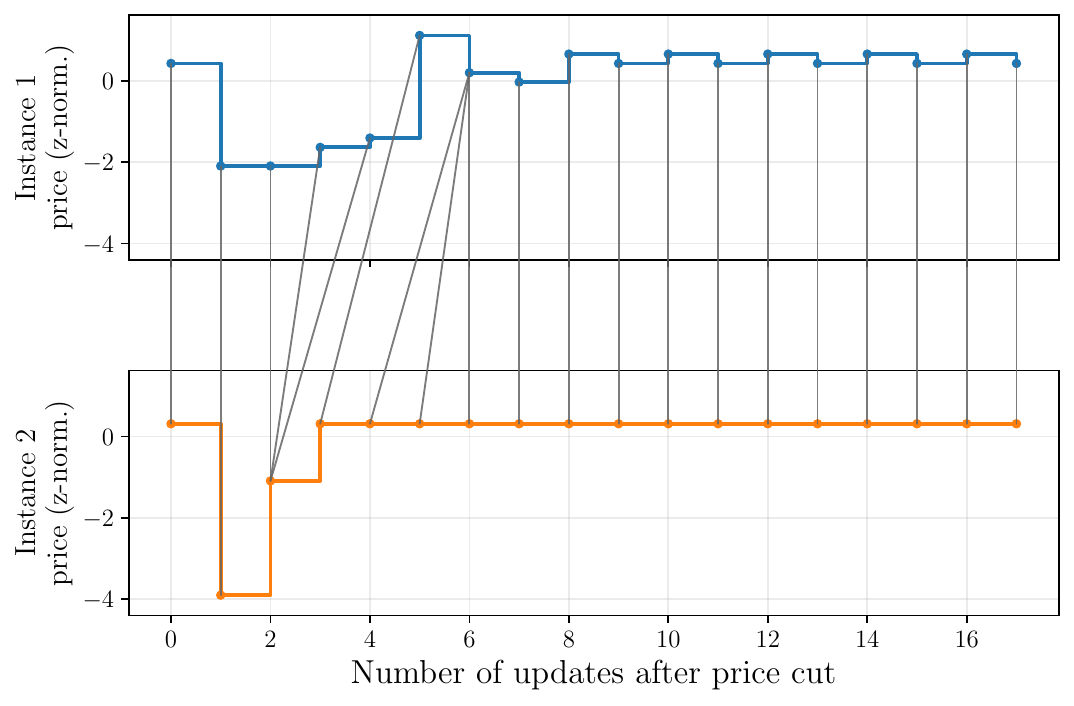}
        \caption{$q=1$, \,$T=10^6$}
        \label{fig:1_10^6}
    \end{subfigure}
    \hfill
    \begin{subfigure}[b]{0.325\textwidth}
        \centering
        \includegraphics[width=\linewidth]{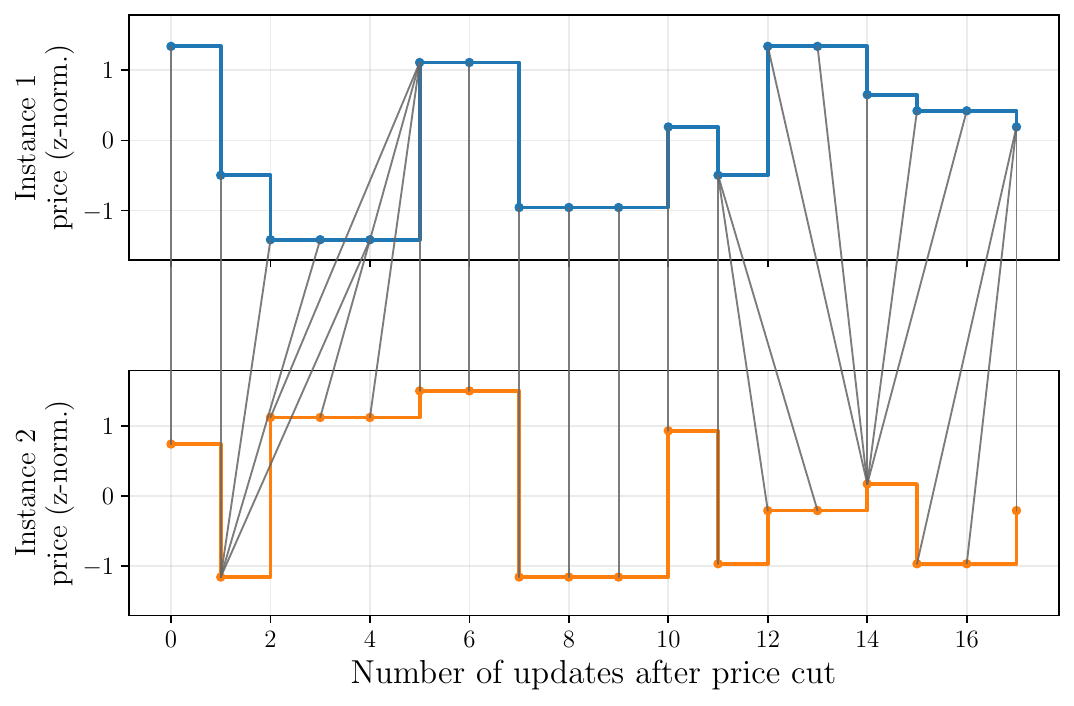}
        \caption{$q=0$, \,$T=10^6$}
        \label{fig:0_10^6}
    \end{subfigure}
    \hfill
    \begin{subfigure}[b]{0.325\textwidth}
        \centering
        \includegraphics[width=\linewidth]{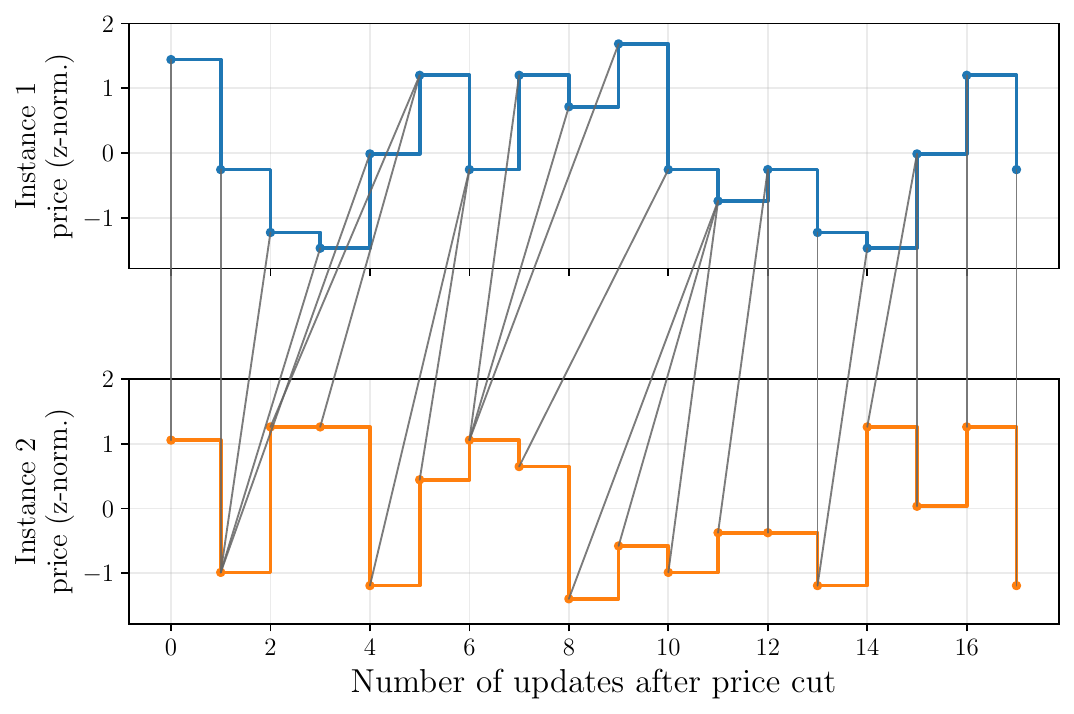}
        \caption{$q=1$, \,$T=1$ (untrained)}
        \label{fig:1_1}
    \end{subfigure}

    \caption{\textbf{Reactions to price cuts and corresponding DTW matches}. Two distinct instances of the \textsc{AvgPrice} specification for (a) algorithms trained with synchronous updates, (b) algorithms trained with asynchronous updates, and (c) untrained algorithms are displayed. The DTW distance allows to compare the \emph{patterns} and finds that the two instances of (a) are similar while there is no clear pattern in (b) and (c).}
    \label{fig:dtw}
\end{figure}

\Cref{fig:dtw} shows DTW's matching output between two instances of reactions to price cuts in three different specifications. \Cref{fig:1_10^6} illustrates how DTW captures the punishment-then-return-to-high-price pattern even though the return is slower in the first instance than in the second. In our experiments, we run $10$ instances of the algorithms per value of $q$ and record the reactions to $10$ price cuts on the final $Q$-matrices with no exploration. We follow the same procedure with $Q$-learning algorithms that are updated only once, so that their behavior is entirely driven by the (randomly chosen) initial conditions. We then proceed to compute the DTW distance between each pair of recorded time-series.

\subsection{Distinguishing trained from untrained reactions}

Given a value of $q$, our goal in this part is to see whether it is possible to distinguish in an unsupervised manner the trajectories of actual $Q$-learning algorithms that have been trained over $T=10^6$ epochs from those that are only driven by initial conditions. 

\begin{figure}
    \centering
    \includegraphics[width=\linewidth]{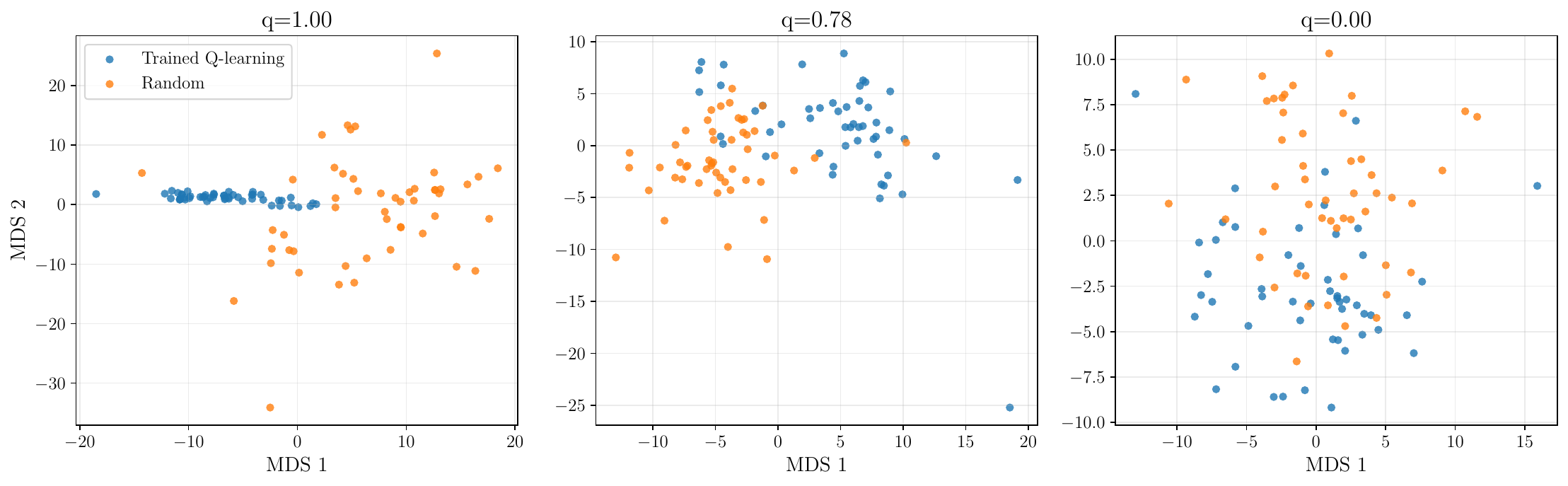}
\caption{\textbf{PCoA of DTW distances in the \textsc{AvgPrice} specification.} Colors represent the ground truth. As $q$ decreases, the two types of points become hardly distinguishable.}
\label{fig: PCoA}
\end{figure}

\Cref{fig: PCoA} shows a projection computed by the classical Multidimensional Scaling (PCoA; \citet{gower1966some}) of the DTW distances on a two-dimensional plane for different values of $q$, in the \textsc{AvgPrice} specification. Visually, we can see for $q=1$ a clear dense cluster of points corresponding to the responses of trained algorithms, while for $q=0$ they are hardly distinguishable from the random ones. To automatically detect the responses of trained algorithms, we use the DBSCAN algorithm \cite{ester1996density}, a tool adapted for the detection of structured clusters among noisy points. It is a clustering algorithm that allows one to detect regions with a high density of points (\textit{clusters}) in a metric space, and distinguish them from noise points. For each data object, it outputs a clustering label, which is the cluster id in which the object has been clustered into, or as a noise point.

One of the advantages of this clustering method is that it does not require the number of clusters as input. We thus do not need to make the assumption that there are only two clusters corresponding to random and $Q$-learning algorithms: the clustering algorithm should detect it on its own. In order to measure the performance of the clustering algorithms in distinguishing the two, the Adjusted Rand Index (ARI) is computed between the clustering labels and the actual class label (\ie random or $Q$-learning). The ARI is a measure of \emph{agreement} between two hard partitions of the data (\ie each data object belongs to only one cluster). 
In our case, an ARI equal to $1$ means that the clustering matches exactly the random vs. $Q$-learning labeling, an ARI score close to $0$ corresponds to a random clustering agreement, while negative ARI values suggest a worse-than-random agreement.

\subsection{Results} \label{sec: results}

\begin{figure}[t]
    \centering

    \begin{subfigure}{0.9\textwidth}
        \centering
        \includegraphics[width=\linewidth]{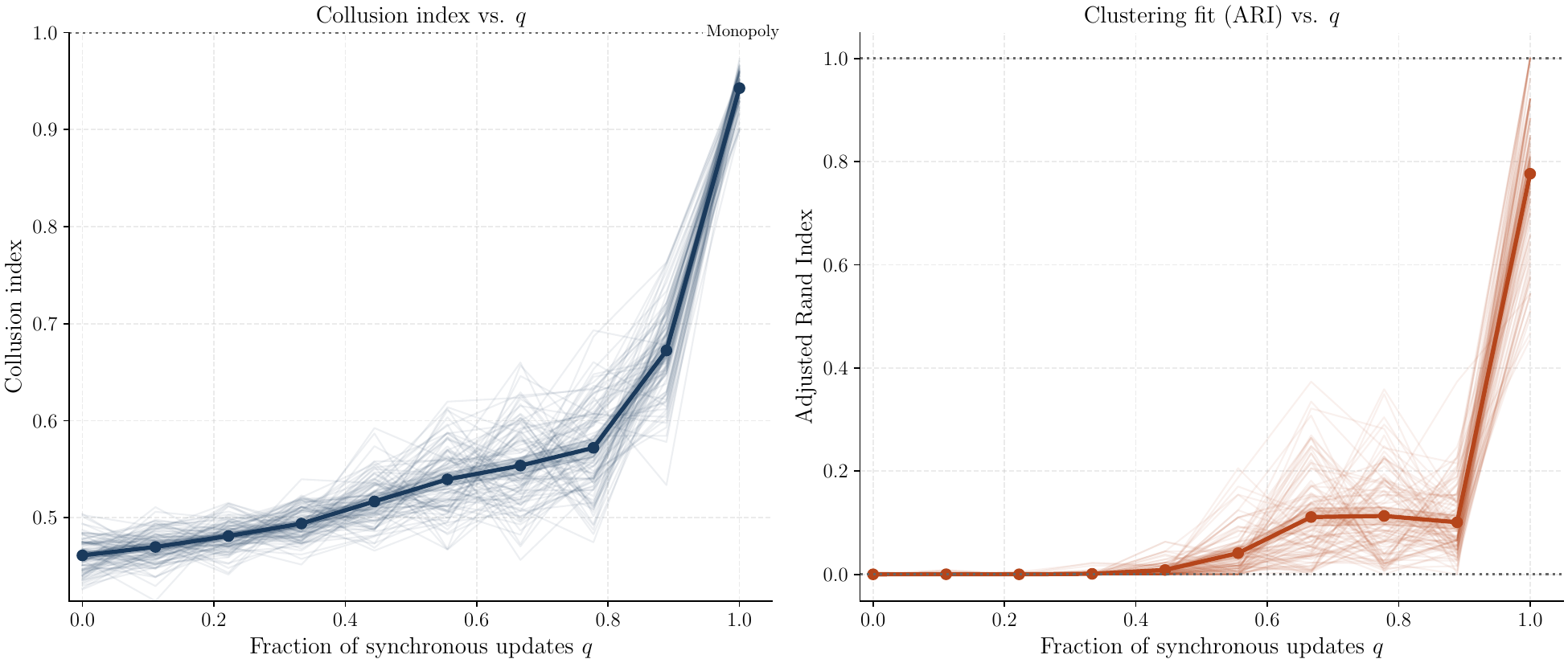}
        \caption{\textsc{AvgPrice}}
        \label{fig:average}
    \end{subfigure}

    \vspace{0.5em}

    \begin{subfigure}{0.9\textwidth}
        \centering
        \includegraphics[width=\linewidth]{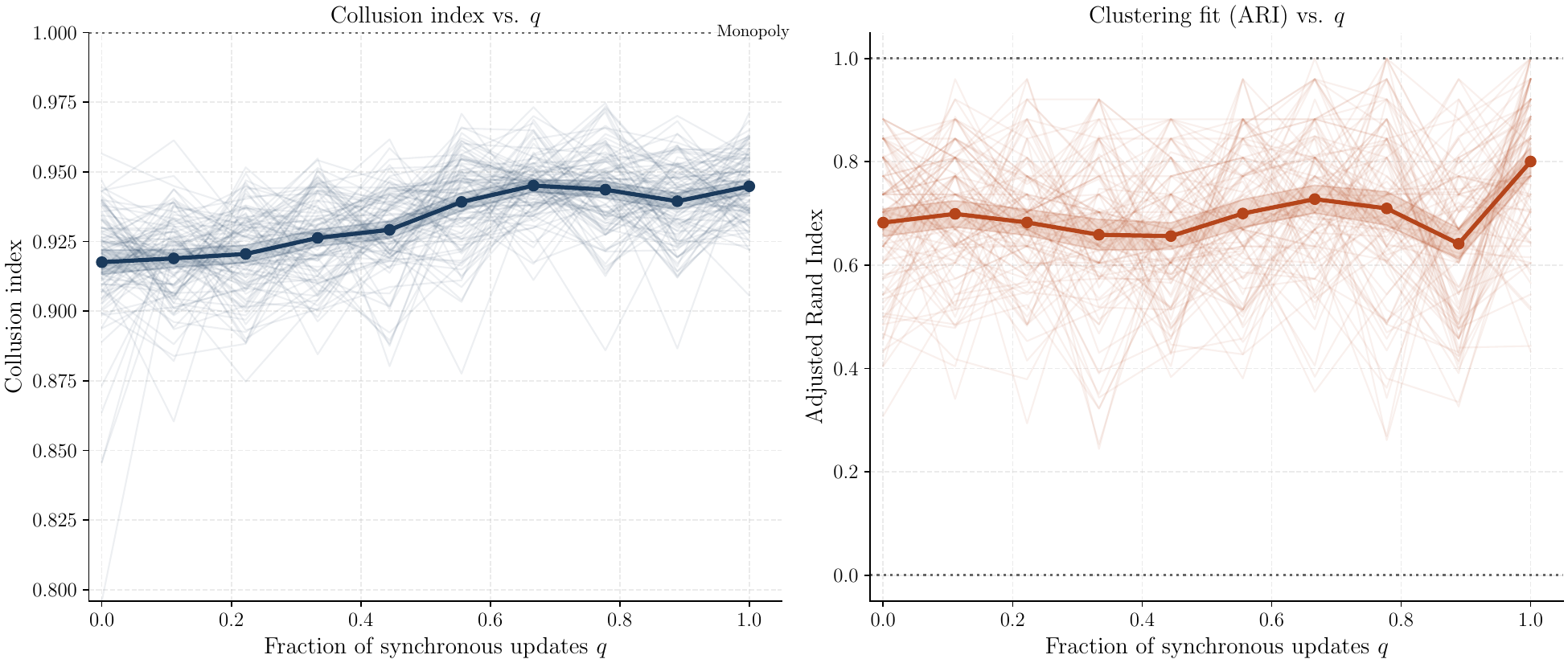}
        \caption{ \textsc{CurrentPrice}}
        \label{fig:current}
    \end{subfigure}

    \caption{\textbf{Collusion index and ARI score} for the \textsc{AvgPrice} (top) and \textsc{CurrentPrice} (bottom) specifications.}
    \label{fig:results}
\end{figure}

\Cref{fig:results} shows the evolution of the collusion index and ARI score for both specifications. The effect of synchrony, captured by $q$, is clear: more synchrony induces more collusion reached by the algorithms. However, there is a stark difference between the two specifications, as qualitatively observed above in \Cref{fig:response_graphs}. When the algorithms condition their price on the current price of their opponent, collusion is much more robust to asynchrony: the collusion index remains very high for $q=0$ and the cluster analysis reveals the consistent presence of reward-punishment strategies. By contrast, when the algorithms condition on their opponent's average price since their own last update, collusion is much more fragile. For low values of $q$, hardly any reward-punishment strategies are detected, and the collusion index is barely above the one attained by algorithms who play randomly.

\section{Discussion}
\label{sec:discussion}
Our results are of immediate relevance for the regulation of algorithmic pricing as they highlight a new channel that affects the appearance of collusion. More specifically, we find that synchrony interacts with the information the algorithms are able to use to condition their action on. A first consequence of our results is to downplay the importance of ``spurious collusion'', that is the collusive outcomes learned by algorithms without conditioning on anything. Our results in \Cref{sec: stateless} show that this type of ``collusion'' disappears without synchronized updates as it precisely relies on a mechanism that needs them. Without a way for the algorithms to synchronize their updates, or an agreement by their designers to do so, which falls under the Sherman Act, this type of collusion does not appear. Whether or not there is a possibility for the algorithms to synchronize their updates strongly depends on the specific environment in which they operate. In a repeated auction, there is a naturally discrete and synchronous timing. However, in a retail market, this is very unlikely: new customers arrive continuously and price updates are often necessary for exogenous reasons such as inventory constraints. 


These findings also indicate that asynchrony protects consumers from algorithmic collusion when there is no information available for the algorithms to condition their action on. Thus, in order to minimize the threat of collusion by pricing algorithms, a regulator should either explicitly forbid algorithms to condition on their opponent's price, or strongly restrict the information available to them. It might be the case that neither of these two solutions is feasible or desirable. The use of pricing algorithms has many benefits for consumers, and enforcing strong restrictions on the type that can be used might make their use overly costly. Similarly, reducing market transparency is harmful to consumers. The two initiatives mentioned in the introduction, FuelWatch and MTS-K, were designed to provide more information to consumers before making a purchase decision. An efficient regulation should thus target how the retailers access this information, for instance by regulating scraping practices. On platforms that can restrict the information pricing algorithms have access to, our results suggest that statistics aggregated over time should be disclosed rather than the current price of competitors in order to prevent algorithmic tacit collusion.

\section{Conclusion} \label{sec: conclusion}
In this article, we have investigated a dimension that was previously overlooked in the algorithmic collusion literature, and affects its appearance: the synchrony of updates. By introducing a model in which synchrony of updates is controllable, we showed that \text{spurious collusion} of stateless algorithms is very sensitive to asynchrony, and thus unlikely in many environments. Then, using extensive numerical experiments, we measured the occurrence of reward-punishment schemes that characterize algorithmic collusion when the algorithms can condition their action on their opponent's price. Notably, we showed that collusion is very sensitive to asynchrony when the algorithms condition their price on the opponent's average price since their last update while it is much more robust when they condition their price on the opponent's current price. These results indicate that an efficient regulation against the threat of algorithmic collusion should target the type of information the algorithms have access to.

\appendix
\section{Interpretations of the timing assumption}\label{sec: timing}
Our assumption over the timing of updates covers at least three different models that are mathematically equivalent. The first is one in which each player's updates are dictated by an independent Poisson clock and whenever an update is prescribed for a player, the other player joins it and updates with probability $q$. This corresponds to standard practices in algorithmic pricing in which the algorithm employed by a firm scrapes its competitors' websites and triggers updates when they do.  In such a case, the parameter $q$ controls for how efficient the algorithms are at scraping their competitor's website in real time.
 
It is equivalent to a model in which there are three independent clocks, one per player and one dictating simultaneous updates. Such a model corresponds to cases in which price updates are triggered by specific events (\eg inventory constraints, surges in website traffic), which can concern only one of the retailers (in the case of inventory constraints) or both of them simultaneously (if an underlying demand shock induces a simultaneous surge in traffic on their website). In this case, the parameter $q$ controls for the share of such events that affect both retailers simultaneously. 

Finally, it is equivalent to a model in which price updates are generated exogenously and a third party decides how to allocate these updates between the two retailers. Such a model fits well cases in which the retailers' price updates go through a platform's API, like on Amazon. The choice of $q$ is then a question of platform design, which decides how synchronously the retailers should update.
\section{$Q$-learning} \label{sec: $Q$-learning}
$Q$-learning is a reinforcement learning principle designed to find optimal solutions to optimization problems in a Markov environment. Formally, denote $S$ a (finite) set of states, $A$ a set of actions and $\pi(s,a)$ the (possibly stochastic) reward obtained in state $s$ after taking action $a$. In each period, an action is taken, a reward is realized and the process moves to the next state with a probability $F(s_{t+1}|s_t, a_t) $. The objective is to learn the best policy, i.e. the one that maximizes $\mathbb{E}\Big{[}\sum_{t=0}^{+\infty}\gamma ^t \pi_t\Big{]}$ with $\gamma \in (0,1)$ a discount rate. $Q$-learning is an iterative method that allows to learn the best policy function without information or hypothesis about the transition function $F$. Classically, the Bellman value function in such a problem writes as follows:
\begin{equation}
    \forall s\in S,\ V(s)=\mathbb{E}(\pi(s,a))+ \gamma \mathbb{E}(V(s')).
\end{equation}
The $Q$-matrix assigns a value to each state-action pair, and is defined as
\begin{equation}
   \forall (s,a)\in A\times S,\  Q(s,a)=\mathbb{E}[\pi | s,a] + \gamma \mathbb{E}[\max_{\{a' \in A\}}Q(s',a') | s, a],
\end{equation}
and linked to the Bellman value function as follows
\begin{equation}
    V(s)=\max_{a\in A}Q(s,a).
\end{equation}
An agent who knows the $Q$-matrix exactly knows what action to take in each state, and thus knows the optimal policy. $Q$-learning estimates this matrix by an iterative procedure. The agent begins with an arbitrary $Q_0$ and updates any cell of the matrix she visits as follows:
\begin{equation} \label{eq: update rule}
    Q_{t+1}(s,a)=(1-\alpha) Q_t(s,a) + \alpha \Big{[} \pi_t + \gamma \max_{a'\in A} Q_t(s',a) \Big{]},
\end{equation}
where $\alpha$ is referred to as the \textit{learning rate} and controls how rapidly $Q$-values change when a reward is obtained. Note that when $\alpha=0$ the $Q$-values stay constant, so that the agent does not learn, and when $\alpha=1$ the $Q$-values immediately  change to the actualized reward. 

Convergence to the optimal Markov policy is guaranteed under conditions on the learning rates and on the exploration policy \citep{singh2000convergence}:
\begin{proposition01}
Given a GLIE policy (Greedy in the Limit with Infinite Exploration), \textit{i.e.} such that:
\begin{itemize}
    \item the exploration policy converges to the greedy one as $t$ goes to $\infty$,
    \item every action-state pair is visited infinitely often,
\end{itemize}
if the sequence of learning rates satisfies:
\begin{enumerate}
    \item $\sum_{t\in \mathbb{N}}\alpha_t=+\infty$,
    \item $\sum_{t\in \mathbb{N}}\alpha_t^2<+\infty$,
\end{enumerate}
then $(Q_t)_{t\in \mathbb{N}}$ converges to the $Q$-matrix with probability $1$.
\end{proposition01}

\section{Experimental setup with states}
For $q\in [0,1]$ we run the algorithms over a horizon $T=\frac{2T_1}{q+1}$ where $T_1=10^6$ and set $\beta= \frac{q+1}{2}\beta_1$ with $\beta_1=10^{-5}$. This ensures that the average number of updates and the final exploration rate remains the same over all the values of $q$. We repeat this process $10$ times to obtain the collusion index and record $5$ reactions to price cuts per run. The reactions to price cuts of each player are recorded over $30$ price updates. The same task is performed for a time horizon of $T_0=1$ for the untrained algorithms. Thus, we obtain a set of $2\times 100$ reactions to price cuts (one per player), $50$ of which are of trained algorithms and $50$ of which are of untrained algorithms.

We then proceed to compute the DTW distances (see \cref{sec: DTW}) between all the recorded reactions to price cuts. More precisely, when simulating the price cut, we exogenously force one of the algorithms to cut its price to a random lower price and record its opponent's reaction as well as its own. For any pair of price cuts, we compute the DTW distance between the reactions of the deviators, the DTW distance between the reactions of the non-deviators, and take the average of the two. Formally, the distance we use is:
\begin{equation}
    D=\frac{1}{2}\text{DTW}(P_{\text{dev}}, P'_{\text{dev}}) + \frac{1}{2}\text{DTW}(P_{-\text{dev}}, P'_{-\text{dev}}),
\end{equation}
where $P_{i}=\{p_{t_k}^i, 0\le k \le30 \}$ and $P'_{i}=\{p_{t_k}^i, 0\le k \le30 \}$ are the prices set by player $i \in \{\text{dev}, -\text{dev}\}$ when they update for the $k$-th time after the price cut.

We then use the distances to perform a clustering task using DBSCAN (see \cref{sec: DBSCAN}). For the minimum neighborhood size, we choose $m=4$ and the radius $\eta$ is set to $10$. The threshold $\eta$ was chosen \textit{a priori} on the scale of the DTW distances, without optimization. Because the average trained-trained distances at $q = 1$ are around $13$ versus $22$ across groups, it is a conservative choice and under-reports the separation between trained and untrained reactions. 

The outcome of the DBSCAN algorithm is a list of labels corresponding to assigned clusters, one for each data point. We compare the assigned clusters to the ground truth by computing a standard ARI index. The ARI reasons over pairs of data points: a pair is concordant if both the DBSCAN partition and the ground-truth partition treat it the same way by either assigning the two points to the same cluster or separating them. Let $a$ be the number of pairs grouped together by both partitions and
$b$ the number separated by both, the Rand index is the fraction of concordant pairs:
\begin{equation}
    RI \;=\; \frac{a+b}{\binom{n}{2}},
\end{equation}
where $n$ is the number of data points. Because this quantity stays high even for unrelated partitions, it is corrected by its expected value under random labeling $\mathbb{E}[RI]$ and normalized:
\begin{equation}
    ARI \;=\; \frac{RI - \mathbb{E}[RI]}{1 - \mathbb{E}[RI]}.
\end{equation}
It equals $1$ when the DBSCAN labels exactly match the ground truth, $0$ when they do not perform better than chance, and can be negative for worse-than-chance agreement.

\section{Dynamic Time Warping} \label{sec: DTW}
Dynamic Time Warping (DTW) is a method for measuring the similarity between two time-series even when these sequences are not perfectly aligned or do not have the same length. The central idea is to find the best possible correspondence between the points of the two series, allowing for local temporal shifts and stretches.

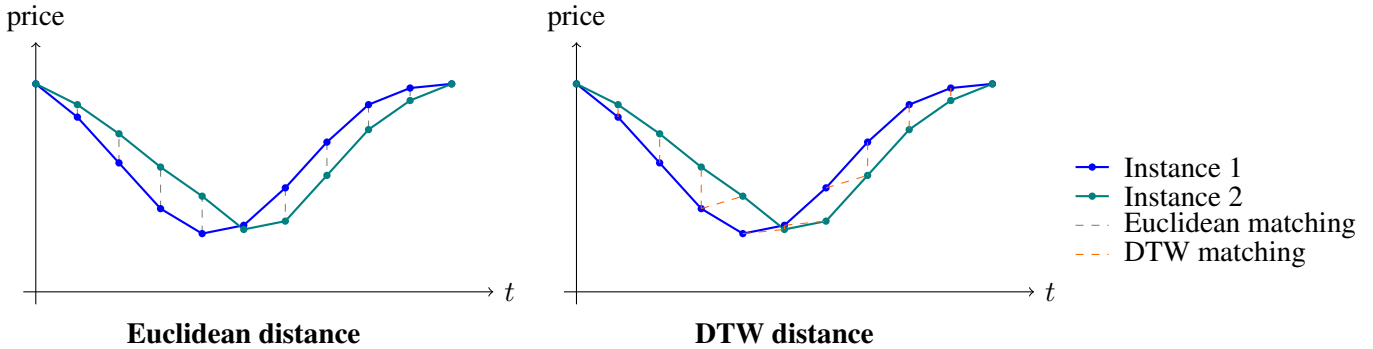
\begin{figure}[h]
\centering
\begin{tikzpicture}[scale=0.55]

\begin{scope}[xshift=0cm]

  \draw[->] (-0.3,0) -- (11,0) node[right] {\small $t$};
  \draw[->] (0,-0.3) -- (0,6) node[above] {\small price};

  \draw[blue, thick]
    (0,5) -- (1,4.2) -- (2,3.1) -- (3,2.0) -- (4,1.4) --
    (5,1.6) -- (6,2.5) -- (7,3.6) -- (8,4.5) -- (9,4.9) -- (10,5.0);

  \draw[teal, thick]
    (0,5) -- (1,4.5) -- (2,3.8) -- (3,3.0) -- (4,2.3) --
    (5,1.5) -- (6,1.7) -- (7,2.8) -- (8,3.9) -- (9,4.6) -- (10,5.0);

  \foreach \x/\ya/\yb in {
    0/5/5, 1/4.2/4.5, 2/3.1/3.8, 3/2.0/3.0, 4/1.4/2.3,
    5/1.6/1.5, 6/2.5/1.7, 7/3.6/2.8, 8/4.5/3.9, 9/4.9/4.6, 10/5.0/5.0
  }{
    \draw[gray, dashed, thin] (\x,\ya) -- (\x,\yb);
    \fill[blue] (\x,\ya) circle (2.5pt);
    \fill[teal] (\x,\yb) circle (2.5pt);
  }

  \node at (5,-1.0) {\small\textbf{Euclidean distance}};

\end{scope}

\begin{scope}[xshift=13cm]

  \draw[->] (-0.3,0) -- (11,0) node[right] {\small $t$};
  \draw[->] (0,-0.3) -- (0,6) node[above] {\small price};

  \draw[blue, thick]
    (0,5) -- (1,4.2) -- (2,3.1) -- (3,2.0) -- (4,1.4) --
    (5,1.6) -- (6,2.5) -- (7,3.6) -- (8,4.5) -- (9,4.9) -- (10,5.0);

  \draw[teal, thick]
    (0,5) -- (1,4.5) -- (2,3.8) -- (3,3.0) -- (4,2.3) --
    (5,1.5) -- (6,1.7) -- (7,2.8) -- (8,3.9) -- (9,4.6) -- (10,5.0);

  \foreach \x/\ya in {
    0/5,1/4.2,2/3.1,3/2.0,4/1.4,
    5/1.6,6/2.5,7/3.6,8/4.5,9/4.9,10/5.0
  }{
    \fill[blue] (\x,\ya) circle (2.5pt);
  }

  \foreach \x/\yb in {
    0/5,1/4.5,2/3.8,3/3.0,4/2.3,
    5/1.5,6/1.7,7/2.8,8/3.9,9/4.6,10/5.0
  }{
    \fill[teal] (\x,\yb) circle (2.5pt);
  }

  \draw[orange!80!red,dashed,thin] (0,5) -- (0,5);
  \draw[orange!80!red,dashed,thin] (1,4.2) -- (1,4.5);
  \draw[orange!80!red,dashed,thin] (2,3.1) -- (2,3.8);
  \draw[orange!80!red,dashed,thin] (3,2.0) -- (3,3.0);
  \draw[orange!80!red,dashed,thin] (3,2.0) -- (4,2.3);
  \draw[orange!80!red,dashed,thin] (4,1.4) -- (5,1.5);
  \draw[orange!80!red,dashed,thin] (5,1.6) -- (5,1.5);
  \draw[orange!80!red,dashed,thin] (5,1.6) -- (6,1.7);
  \draw[orange!80!red,dashed,thin] (6,2.5) -- (7,2.8);
  \draw[orange!80!red,dashed,thin] (7,3.6) -- (7,2.8);
  \draw[orange!80!red,dashed,thin] (8,4.5) -- (8,3.9);
  \draw[orange!80!red,dashed,thin] (9,4.9) -- (9,4.6);
  \draw[orange!80!red,dashed,thin] (10,5.0) -- (10,5.0);

  \node at (5,-1.0) {\small\textbf{DTW distance}};

\end{scope}

\begin{scope}[xshift=25cm,yshift=3cm]

  \draw[blue,thick] (0,0) -- (0.8,0);
  \fill[blue] (0.4,0) circle (2.5pt);
  \node[right] at (0.9,0) {\small Instance 1};

  \draw[teal,thick] (0,-0.7) -- (0.8,-0.7);
  \fill[teal] (0.4,-0.7) circle (2.5pt);
  \node[right] at (0.9,-0.7) {\small Instance 2};

  \draw[gray,dashed,thin] (0,-1.4) -- (0.8,-1.4);
  \node[right] at (0.9,-1.4) {\small Euclidean matching};

  \draw[orange!80!red,dashed,thin] (0,-2.1) -- (0.8,-2.1);
  \node[right] at (0.9,-2.1) {\small DTW matching};

\end{scope}

\end{tikzpicture}
\caption{Comparison of Euclidean (point-to-point) and DTW distance for two reaction profiles sharing the same reward-punishment structure but differing in the speed of adjustment.}
\label{fig: comparison dtw euclidean}
\end{figure}

To build intuition for the problem DTW solves, consider two synchronous algorithms that have learned a reward-punishment strategy. Following a unilateral price cut, both algorithms punish the deviating firm by lowering their own price before reverting to the collusive level. However, the precise timing of these adjustments may differ across algorithms, depending on the learning phase. A point-by-point comparison of their reaction profiles would penalize these minor temporal discrepancies and incorrectly conclude that the two algorithms have different behaviors. In contrast, DTW allows for a variable temporal offset and correctly identifies the structural similarity of the two reward-punishment patterns. \Cref{fig: comparison dtw euclidean} illustrates the advantage of DTW over the euclidean distance.

\subsection{Formalization}

Let $A = (a_1, a_2, \ldots, a_m)$ and $B = (b_1, b_2, \ldots, b_n)$ be two time-series of lengths $m$ and $n$ respectively. An \emph{alignment path} $W = (w_1, w_2, \ldots, w_K)$ is a sequence of index pairs, where each element $w_k = (i, j)$ indicates that point $a_i$ is matched to point $b_j$. This path must satisfy three constraints.

\medskip
\noindent\textbf{Boundary condition.} The path begins at $(1,1)$ and ends at $(m,n)$: the starts and ends of both series are necessarily matched to one another.

\medskip
\noindent\textbf{Monotonicity.} If $w_k = (i, j)$ and $w_{k+1} = (i', j')$, then $i' \geq i$ and $j' \geq j$: the path cannot move backward in time.

\medskip
\noindent\textbf{Continuity.} The path advances by at most one step at a time: $i' - i \leq 1$ and $j' - j \leq 1$.

\medskip
The DTW distance between $A$ and $B$ is then defined as the minimum total alignment cost over all admissible paths:
\begin{equation}
    \mathrm{DTW}(A, B) = \min_W \sum_{k=1}^{K} d(a_{i_k}, b_{j_k})
\end{equation}
where $d(a_i, b_j)$ is a local distance between points $a_i$ and $b_j$ --- typically the Euclidean distance $|a_i - b_j|$.

\subsection{Computation via dynamic programming}

The DTW distance is computed efficiently via \emph{dynamic programming}. One builds a cumulative cost matrix $D$ of size $m \times n$, defined by the recurrence:
\begin{equation}
    D(i, j) = d(a_i, b_j) + \min\bigl(D(i-1, j),\; D(i, j-1),\; D(i-1, j-1)\bigr)
\end{equation}
with the initial condition $D(1,1) = d(a_1, b_1)$. The DTW distance is then read off as $D(m,n)$. The computational complexity of this procedure is $\mathcal{O}(mn)$ in both time and space.

\section{DBSCAN} \label{sec: DBSCAN}
Once pairwise DTW distances between reaction profiles have been computed, we use the
DBSCAN algorithm \citep{ester1996density} (\textit{Density-Based Spatial Clustering
of Applications with Noise}) to partition these profiles into groups of similar
reactions. Unlike $k$-means, DBSCAN does not require the number of clusters to be
specified in advance, and is specifically designed to distinguish clustered points from noise points: this is precisely our goal, as we aim to distinguish the patterns learned by actual $Q$-learning algorithms from the ones of random algorithms. The algorithm operates as follows.


DBSCAN takes two hyperparameters as input: a radius $\eta > 0$ and a minimum number of points $m \geq 1$. Given a set of $n$ reaction profiles and their pairwise DTW distances, the algorithm first classifies each profile as a \textit{core point}, a \textit{border point}, or a \textit{noise point}. A profile $p$ is a core point if at least $m$ other profiles lie within DTW distance $\eta$ of $p$, that is, if $p$ has a sufficiently dense neighborhood. A profile $q$ is a border point if it lies within distance $\eta$ of some core point but does not itself have a dense neighborhood. All remaining profiles are classified as noise points and are left unassigned to any cluster. 

Clusters are then formed by a simple connectivity rule: two core points belong to the same cluster if and only if their DTW distance is at most $\eta$. Border points are assigned to the cluster of the core point they are closest to. This process yields a partition of the non-noise profiles into clusters, where each cluster can be thought of as a dense region of the DTW distance space, separated from other clusters by regions of lower density. \Cref{alg:dbscan} implements DBSCAN using the DTW metric.

\bigskip

\begin{algorithm}[H]\small
\caption{DBSCAN with DTW distance}
\label{alg:dbscan}
\begin{algorithmic}[1]
\Require Reaction profiles $\mathcal{P} = \{p_1, \ldots, p_n\}$, 
         radius $\eta > 0$, 
         minimum neighborhood size $m \geq 1$
\Ensure  Cluster labels $\ell_1, \ldots, \ell_n$ 
         (with $\ell_i = \textsc{noise}$ if $p_i$ is unassigned)

\medskip
\State Initialize all labels: $\ell_i \leftarrow \textsc{unvisited}$ for all $i$
\State $c \leftarrow 0$ \Comment{Cluster counter}

\medskip
\For{each profile $p_i \in \mathcal{P}$}
    \If{$\ell_i \neq \textsc{unvisited}$}
        \State \textbf{continue}
    \EndIf
    \State $\mathcal{N}(p_i) \leftarrow \{p_j \in \mathcal{P} : \mathrm{DTW}(p_i, p_j) \leq \eta\}$
    \If{$|\mathcal{N}(p_i)| < m$}
        \State $\ell_i \leftarrow \textsc{noise}$ \Comment{$p_i$ is a noise point}
    \Else
        \State $c \leftarrow c + 1$
        \State $\ell_i \leftarrow c$ \Comment{$p_i$ is a core point; start new cluster $c$}
        \State $\mathcal{S} \leftarrow \mathcal{N}(p_i) \setminus \{p_i\}$ 
               \Comment{Seeds: neighbors to expand}
        \While{$\mathcal{S} \neq \emptyset$}
            \State Pick any $p_j \in \mathcal{S}$; remove $p_j$ from $\mathcal{S}$
            \If{$\ell_j = \textsc{noise}$}
                \State $\ell_j \leftarrow c$ 
                       \Comment{$p_j$ is a border point of cluster $c$}
            \EndIf
            \If{$\ell_j = \textsc{unvisited}$}
                \State $\ell_j \leftarrow c$
                \State $\mathcal{N}(p_j) \leftarrow 
                       \{p_k \in \mathcal{P} : \mathrm{DTW}(p_j, p_k) \leq \eta\}$
                \If{$|\mathcal{N}(p_j)| \geq m$}
                    \State $\mathcal{S} \leftarrow \mathcal{S} \cup \mathcal{N}(p_j)$
                           \Comment{$p_j$ is also a core point; expand further}
                \EndIf
            \EndIf
        \EndWhile
    \EndIf
\EndFor
\end{algorithmic}
\end{algorithm}
\newpage
\section*{Acknowledgements}
This work has benefited from the support of the Agence Nationale de la Recherche through the program Investissements d’Avenir ANR-17-EURE-0001. The project leading to this publication has received funding from the French government under the “France 2030” investment plan managed by the French National Research Agency (reference:ANR-17-EURE-0020) and from Excellence Initiative of Aix-Marseille University - A*MIDEX.
Argyris Kalogeratos acknowledges support by the Industrial Analytics and Machine Learning (IdAML)
Chair hosted at ENS Paris-Saclay, University Paris-Saclay. 

\newpage
\bibliographystyle{plainnat}
\bibliography{biblio}
\end{document}